\pdfoutput=1

\documentclass[%
aip, apl, numerical,
amsmath,amssymb,
reprint,
graphicx]{revtex4-2}

\usepackage{hyperref}
\hypersetup{colorlinks=true, citecolor=blue, urlcolor=blue, linkcolor=blue}
\usepackage[utf8]{inputenc}
\usepackage{graphicx}
\usepackage{subcaption}
\usepackage{siunitx}
\usepackage{xargs} 
\usepackage{soul}
\usepackage[capitalize,noabbrev]{cleveref}
\usepackage{xcolor}

\usepackage{booktabs}

\newcommandx{\betastr}{\beta^*}

\newcommandx{\kaist}{KAIST}
\newcommandx{\idest}{i.e.\ }
\newcommandx{\plchldr}{\mathord{\color{black!100}\bullet}}%

\DeclareSIUnit{\dBm}{\decibel\milli\relax}

\begin{document}


\title{Characterization of a flux-driven Josephson parametric amplifier with
near quantum-limited added noise for axion search experiments} 


\author{\c{C}a\u{g}lar Kutlu}
\email[]{caglar.kutlu@gmail.com}
\affiliation{Korea Advanced Institute of Science and Technology, Daejeon 34051,
Republic of Korea}
\affiliation{Center for Axion and Precision Physics Research, Institute for
Basic Science, Daejeon 34051, Republic of Korea}
\author{Arjan F. van Loo}
\affiliation{Center for Emergent Matter Science (CEMS), RIKEN, Wako,
Saitama 351--0198, Japan}
\author{Sergey V. Uchaikin}
\author{Andrei N. Matlashov}
\affiliation{Center for Axion and Precision Physics Research, Institute for
Basic Science, Daejeon 34051, Republic of Korea}
\author{Doyu~Lee}
\thanks{Present address: Samsung Electronics, Gyeonggi--do 16677, Republic of Korea}
\affiliation{Center for Axion and Precision Physics Research, Institute for
Basic Science, Daejeon 34051, Republic of Korea}
\author{Seonjeong~Oh}
\author{Jinsu~Kim}
\affiliation{Korea Advanced Institute of Science and Technology, Daejeon 34051,
Republic of Korea}
\affiliation{Center for Axion and Precision Physics Research, Institute for
Basic Science, Daejeon 34051, Republic of Korea}
\author{Woohyun Chung}
\affiliation{Center for Axion and Precision Physics Research, Institute for
Basic Science, Daejeon 34051, Republic of Korea}
\author{Yasunobu Nakamura}
\affiliation{Center for Emergent Matter Science (CEMS), RIKEN, Wako, Saitama 351--0198, Japan}
\affiliation{Research Center for Advanced Science and Technology (RCAST), The University of Tokyo, Meguro--ku, Tokyo 153--8904, Japan}
\author{Yannis K. Semertzidis}
\affiliation{Korea Advanced Institute of Science and Technology, Daejeon 34051, Republic of Korea}
\affiliation{Center for Axion and Precision Physics Research, Institute for
Basic Science, Daejeon 34051, Republic of Korea}


\date{\today}

\begin{abstract}
The axion, a hypothetical elementary pseudoscalar, is expected to solve the
strong \emph{CP} problem of QCD and is also a promising candidate for dark
matter.  The most sensitive axion search experiments operate at millikelvin
temperatures and hence rely on instrumentation that carries signals from a system
at cryogenic temperatures to room temperature instrumentation.  One of the
biggest limiting factors affecting the parameter scanning speed of these detectors is
the noise added by the components in the signal detection chain.  Since the first
amplifier in the chain limits the minimum noise, low-noise amplification is of
paramount importance. This paper reports on the operation of a flux-driven
Josephson parametric amplifier (JPA) operating at around \SI{2.3}{\giga\hertz}
with added noise approaching the quantum limit.  The JPA was employed as a first
stage amplifier in an experimental setting similar to the ones used in
haloscope axion detectors.  By operating the JPA at a gain of \SI{19}{\decibel}
and cascading it with two cryogenic amplifiers operating at \SI{4}{\kelvin}, noise
temperatures as low as \SI{120}{\milli\kelvin} were achieved for the whole signal detection chain.
\end{abstract}






\maketitle 

\section{Introduction}\label{sec:intro}
Axions are spin-0 particles that emerge as a result of the Peccei-Quinn
mechanism which was originally proposed as a solution to the strong \emph{CP} problem
of quantum chromodynamics\cite{Peccei1977,Weinberg1978}.  They were also identified as
viable candidates for all or a fraction of the cold dark matter in
our universe\cite{Preskill1983,Dine1983,Abbott1983}.  It is possible to detect
axions upon their conversion to microwave photons, using resonant cavities
immersed in high magnetic fields\cite{Sikivie1983}.  Since the axion mass is unknown,
these detectors employ a mechanism to scan different frequencies corresponding
to different axion masses.  The scanning rate of such detectors scales with $
1/T_{\mathrm{sys}}^2
$, where $ T_{\mathrm{sys}} $ is the system noise background characterized in units of
temperature.  It can be decomposed as $ T_{\mathrm{sys}} = T_{\mathrm{cav}} +
T_{\mathrm{add}} $, where the first term denotes the noise temperature accompanying
the signal itself and the second one denotes the noise added by the signal
detection chain. Throughout this work, noise temperature refers to the added
noise unless otherwise stated. In order to reduce $ T_{\mathrm{cav}} $, the
cavity is cooled to millikelvin temperatures.  If the first amplifier has
sufficiently high gain ($ G_1 $), its noise temperature ($ T_1 $) will be the
dominant contribution to $ T_{\mathrm{add}} $ as given by the well-known
relation\cite{Friis1944}: $ T_{\mathrm{add}} = T_1 +
\frac{T_\mathrm{rest}}{G_1}$ where  $ T_{\mathrm{rest}} $ is the noise
temperature of the whole chain except the first amplifier.  Amplifiers based on
Josephson junctions including microstrip superconducting quantum interference
device amplifiers (MSA) and JPA have already been shown to be capable of gains
higher than \SI{30}{\decibel}, and noise temperatures approaching the quantum
limit\cite{Castellanos-Beltran2007, Kinion2011}.  While an MSA has an internal
shunt resistor used for biasing which hinders noise
performance\cite{Uchaikin2019, Andre1999}, by design the JPA requires no
resistive element to operate.  Several experiments presently searching for dark
matter axions have already adopted the JPA as the first
amplifier\cite{Braine2020,BrubakerAxion2017,Crescini2020}.  In this work, the
frequency coverage, gain and noise properties of a flux-driven JPA for use in
an axion dark matter experiment operating around \SI{2.3}{\giga\hertz} are
investigated.

The power spectral density of the noise accompanying a signal measured in an impedance
matched environment can be given as~\cite{Callen1951}~:
\begin{align}
\label{eq:noisepow}
S_n(f,T) = hf\left[
	\frac{1}{\exp{\left(\frac{hf}{k_B T}\right)} - 1} + 
    \frac{1}{2}\right]
\end{align}
where $ h $ is Planck's constant and $ k_B $ is Boltzmann's constant.
The first term in the brackets is the mean number of quanta at frequency $ f $
at the bath temperature $ T $ and the second term is the contribution from
zero-point fluctuations.  The lower limit on noise temperature for linear
phase-insensitive amplifiers is given by\cite{Clerk2010} $ T_Q = \lim_{T\rightarrow 0}
S_n(f, T)/(k_B)= hf/(2k_B) $ which is about \SI{55.2}{\milli\kelvin} at \SI{2.3}{\giga\hertz}.   
Using a \SI{2.3}{\giga\hertz} flux-driven JPA $ T_{\mathrm{add}} \approx
\SI{120}{\milli\kelvin}$ is achieved.  This corresponds to a $ T_\mathrm{sys}
\approx \SI{190}{\milli\kelvin} $ for an axion haloscope experiment running at
a bath temperature of \SI{50}{\milli\kelvin}.  The lower bound for $
T_\mathrm{sys} $ is given by the standard quantum limit\cite{Caves1982} $
T_\mathrm{SQL} = 2 T_Q $ which is about \SI{110}{\milli\kelvin} at
\SI{2.3}{\giga\hertz}. 

\section{Flux-driven JPA}
\begin{figure}[h]
    \centering
    \includegraphics[width=0.8\linewidth]{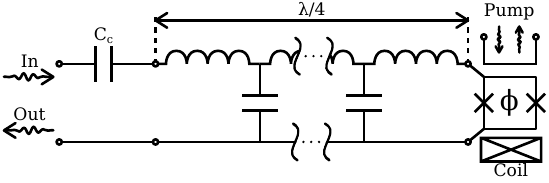}
    \caption{Equivalent circuit diagram of the JPA sample.  The JPA was
        fabricated by photolithography of a Nb layer, deposited on a
        \SI{0.3}{\milli\meter} thick Si substrate.  The SQUID was placed
        on top of the Nb layer by E-beam lithography followed by shadow 
        evaporation\cite{Dolan1977,Zhong2013}. The sample was attached to a printed circuit board (PCB)
        and the transmission lines were bonded with Al wires.  The PCB was fixed 
        onto a gold plated copper structure and placed inside a 
        superconducting coil.  The whole structure was covered tightly with a
        lead shield and attached to the mixing-chamber (MC) plate using a gold plated
        copper rod.}\label{fig:circuit}
\end{figure}

The equivalent circuit diagram of the tested device is shown in
\cref{fig:circuit}. It consists of a superconducting quantum interference
device (SQUID) attached to the end of a coplanar waveguide $ \lambda/4 $
resonator that is coupled via a capacitor ($C_c$) to the 
transmission line for the signal input and output. The SQUID acts as a variable
inductor whose value depends on the magnetic flux passing through its loop.  In
the setup, a superconducting coil is used to provide the necessary DC flux ($\phi$) 
through the SQUID loop in order to tune the resonance frequency
($f_r$).  Parametric amplification is achieved by modulating the flux through
the SQUID using a pump signal.  The pump tone is provided by a separate
transmission line inductively coupled to the SQUID\@.  The JPA is operated in
the three-wave mixing mode\cite{Roy2016} where the pump ($f_p$), the signal
($f_s$), and the idler ($f_i$) frequencies satisfy the relation $ f_p = f_s +
f_i $.  The signal input and output share the same port.  A circulator is used
to separate them.  Since the $ \lambda/4 $ resonator only allows odd harmonics,
there is no measurable pump leakage to the output line.  This prevents the
stronger pump tone from saturating the rest of the amplifiers in the
chain\cite{Yamamoto2008}.  \Cref{fig:setup} shows a schematic for the axion
search experimental setup.

\section{Measurements}
\begin{figure}[h]
    \centering
    \includegraphics[width=0.8\linewidth]{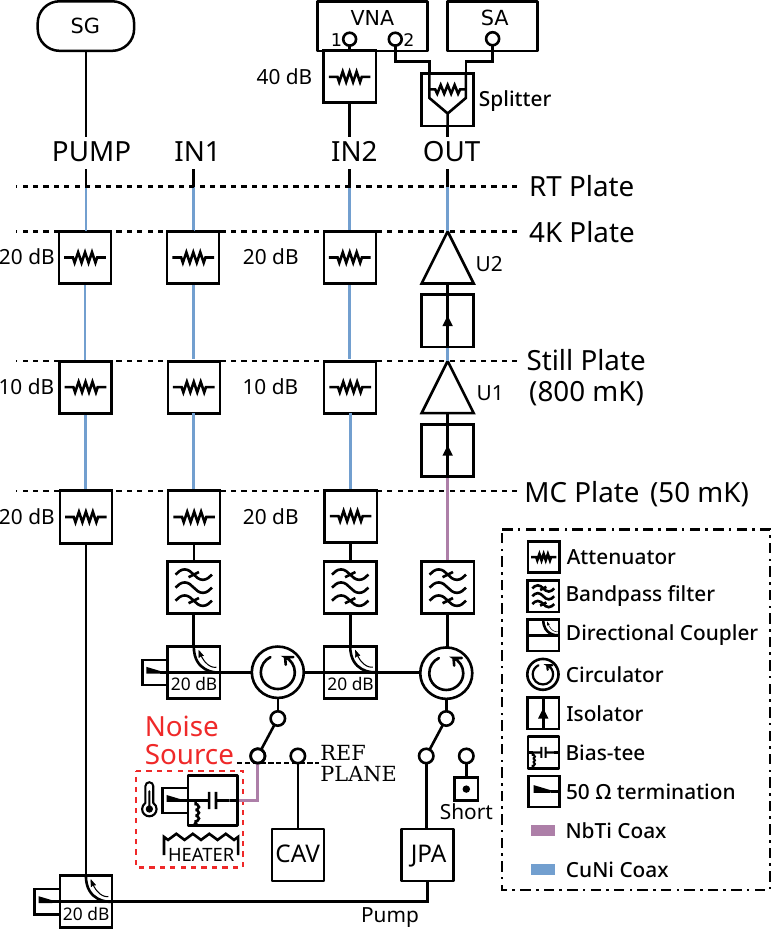}
    \caption{The experimental setup used in all the characterization measurements.
        SG, VNA, and SA stand for the signal generator, vector network analyzer, and
        spectrum analyzer, respectively.  During this work, the switch that
        selects between the cavity and the noise source was always kept at the
        position shown in the figure.  The ports IN2 and OUT were used to
        directly measure the JPA characteristics, bypassing the cavity.  The
        microwave short element shown next to the JPA was used to bypass the
        JPA for calibration measurements.  U1 and U2 are HEMT amplifiers with
        noise temperatures of \SI{1.5}{\kelvin} and \SI{5}{\kelvin},
        respectively.}\label{fig:setup}
\end{figure}

\begin{figure}[h]
    \centering
    \includegraphics[width=0.95\linewidth]{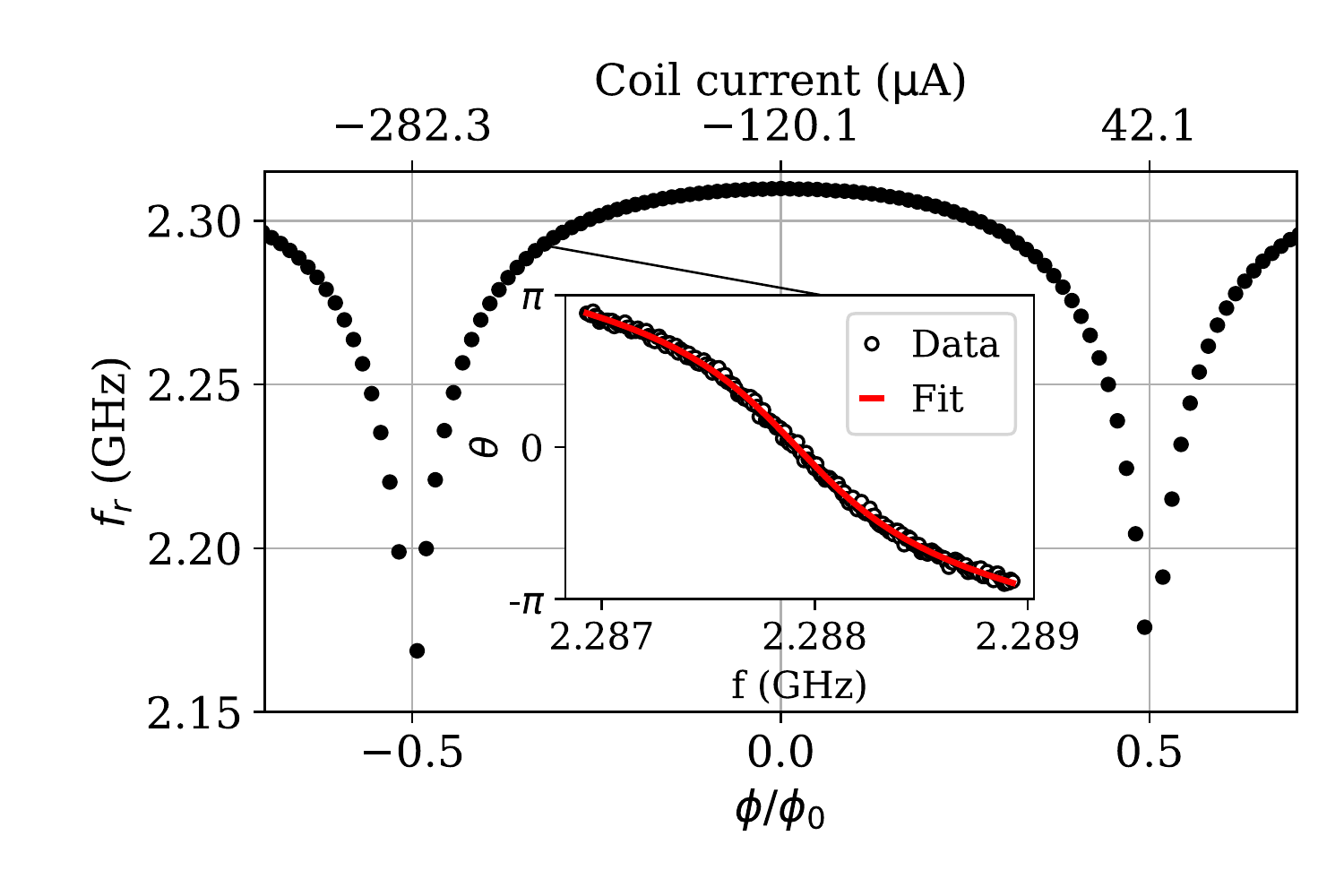}
    \caption{Resonance frequency versus flux obtained by sweeping
        the coil current and measuring the phase response at each step.  One
        period corresponds to a current of \SI{324.4}{\micro\ampere}.  The inset
        shows the fit performed to estimate the resonance frequency for each applied
        flux.}\label{fig:resmap}
\end{figure}

When there is no pump tone present, the JPA can be modeled as a
resonator with a well-defined quality factor and resonance frequency which are
functions of flux.  
%
The resonance frequency is estimated from the frequency domain phase response 
using a parameter fit\cite{Krantz2013}.
The phase response is obtained
by doing a transmission S-parameter measurement using a vector network analyzer
(VNA) in the configuration as shown in \cref{fig:setup}.  The resonance
frequency was measured as a function of the coil current
(see \cref{fig:resmap}).  It was found that the minimum observable resonance frequency was 
at \SI{2.18}{\giga\hertz} and the maximum was \SI{2.309}{\giga\hertz}.
The lower bound is due to the frequency band of the circulators which spans 
from \SI{2.15}{\giga\hertz} to \SI{2.60}{\giga\hertz}. 
At the lower frequencies, the JPA becomes much more sensitive to flux noise due
to a higher $ \frac{\partial f_r}{\partial \phi} $.  This work mainly focused
on operation with frequencies above \SI{2.2}{\giga\hertz}.

During the experiments, the MC plate temperature was stabilized at
\SI[separate-uncertainty = true]{50 \pm{}1}{\milli\kelvin}.  With the
temperature fixed, the frequency response of the JPA is determined by three
experimental variables: the coil current ($ i_b $), the pump
frequency ($ f_p $), and the pump power ($ P_p $). The
measurements shown in this work had $ i_b $ confined to the region where the
flux through the SQUID loop is given by $ -0.5 \phi_0 < \phi < 0 $, where $ \phi_0 $ is
the magnetic flux quantum.  Therefore, $ f_r $ can be unambiguously
converted to $ \phi $ or $ i_b $.  All experiments began with a
transmission measurement, with the resonance
frequency tuned to \SI{2.18}{\giga\hertz}.  This becomes the baseline
measurement to be used for the duration of the experiment.
When the result was compared to a separate measurement, in which a microwave short
was put in place of the JPA, it was found that the baseline obtained via such an
off-resonance measurement was at most \SI{0.2}{\decibel} lossier than an ideal
mirror.  The JPA gain ($ G_J $) was estimated by dividing the transmission magnitude
response with the baseline's magnitude response.

\begin{figure} \centering 
\begin{subfigure}{\linewidth}
    \caption{}
    \includegraphics[width=\linewidth]{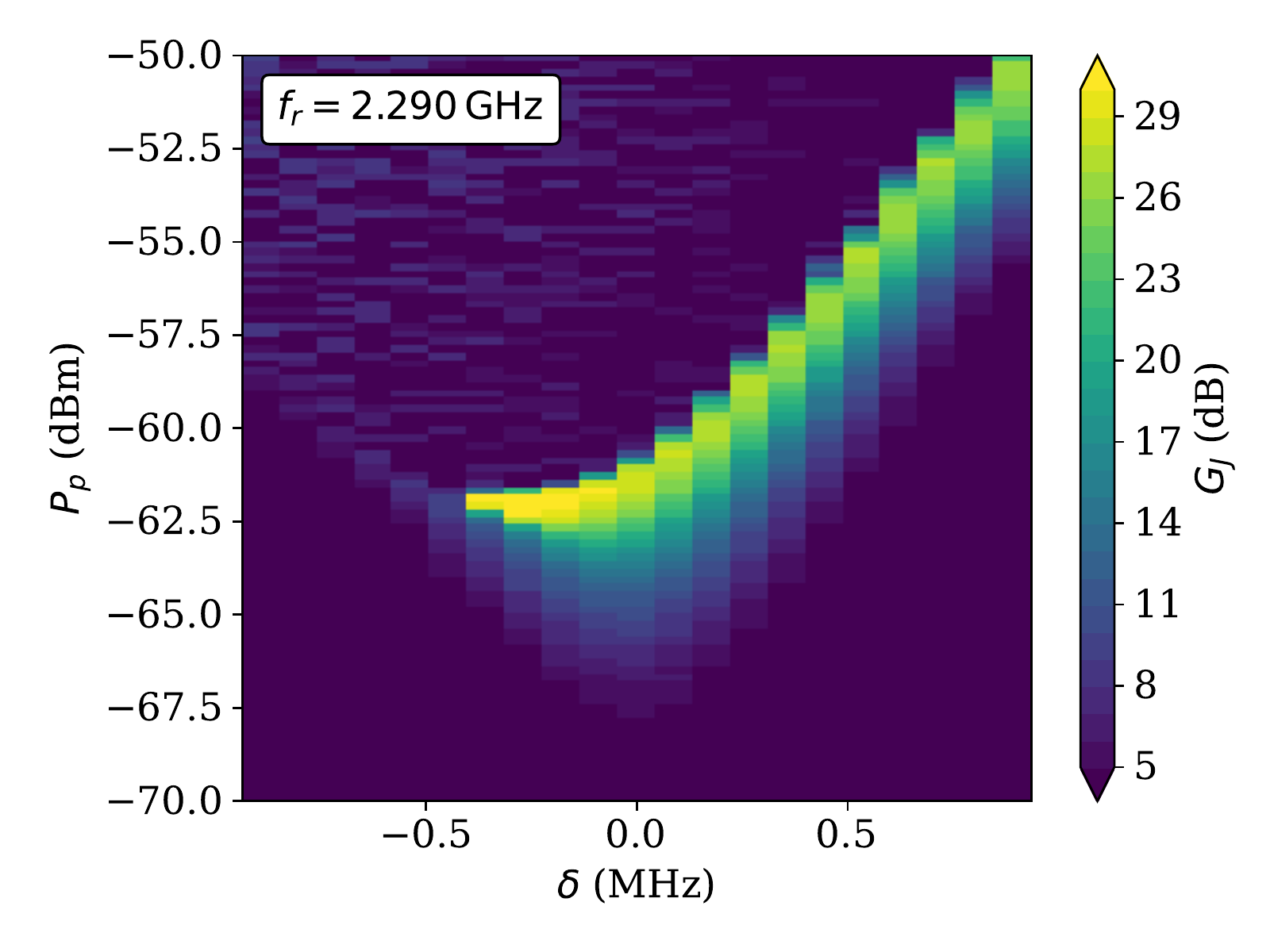}\label{fig:paramap}
\end{subfigure} 
\begin{subfigure}{\linewidth}
    \caption{}
    \includegraphics[width=\linewidth]{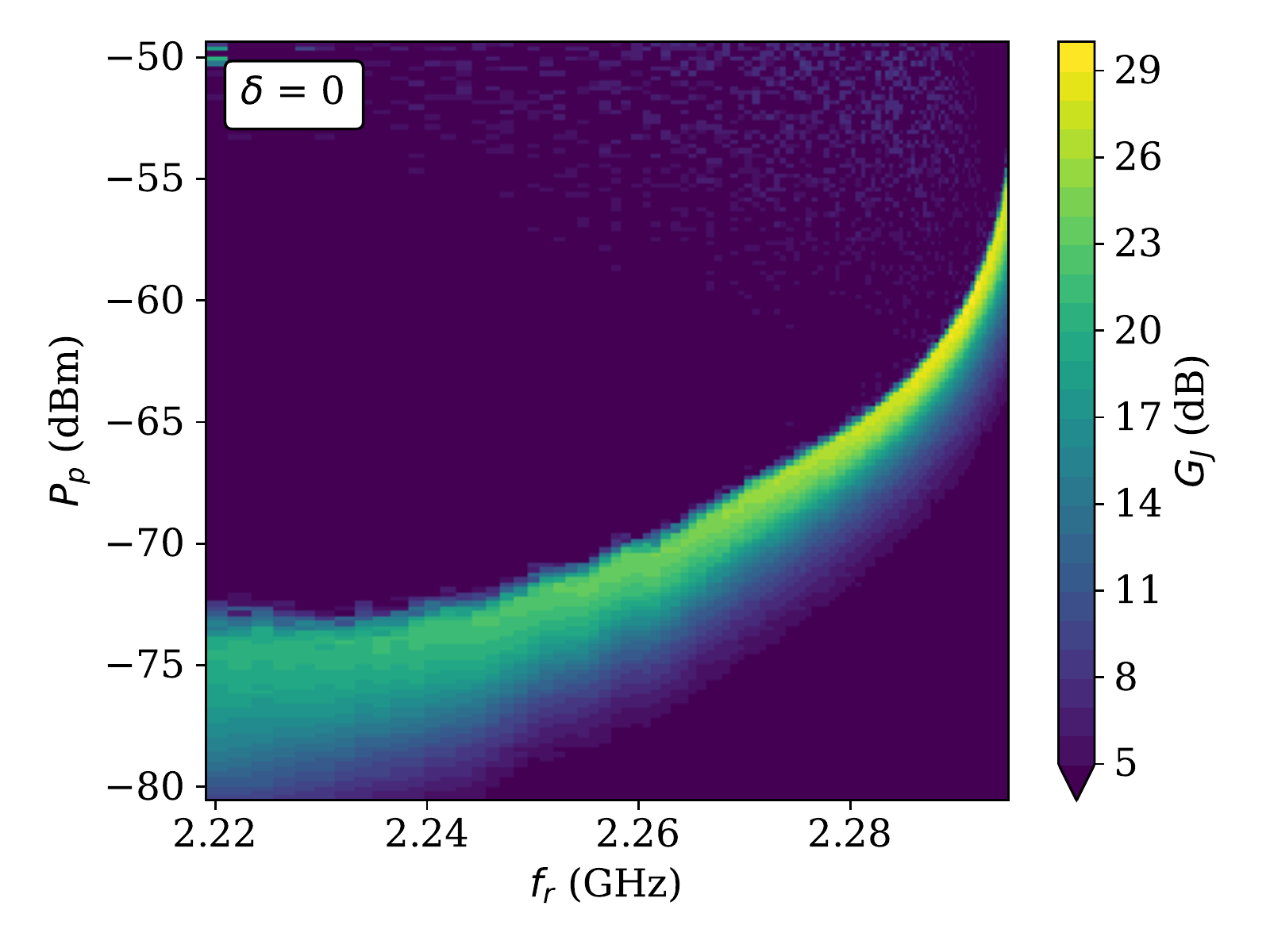}\label{fig:gainsweep}
\end{subfigure}
\caption{(a)~Maximum gain measured as a function of detuning and pump power for
    a flux bias corresponding to $ f_r = \SI{2.29}{\giga\hertz}$.
    (b)~Maximum gain as a function of frequency and pump power with $ f_p = 2 f_r $.}\label{fig:gains}
\end{figure}

To investigate the gain behavior, a sweep over the parameters $ i_b $, $ f_p $,
$ P_p $ was made and the maximum gain was measured at each point.
After each $ i_b $ tuning step, the resonance frequency is estimated by
performing a phase measurement and applying a parameter fit.  With
the detuning defined as $ \delta = f_p/2 - f_r $, the equigain contours had a
minimum in necessary pump power around $ \delta = 0 $, as shown in
\cref{fig:paramap}.  It was observed that for resonance frequencies above
\SI{2.299}{\giga\hertz} the minimum starts to shift to lower detunings which is
attributed to pump-induced shifts in resonance frequency\cite{Krantz2013}.
\Cref{fig:gainsweep} shows that the slice of $ \delta = 0 $ can be used to
achieve peak gains of up to \SI{30}{\decibel} along the frequency range of the
device.

\begin{figure} 
    \centering
    \includegraphics[width=\linewidth]{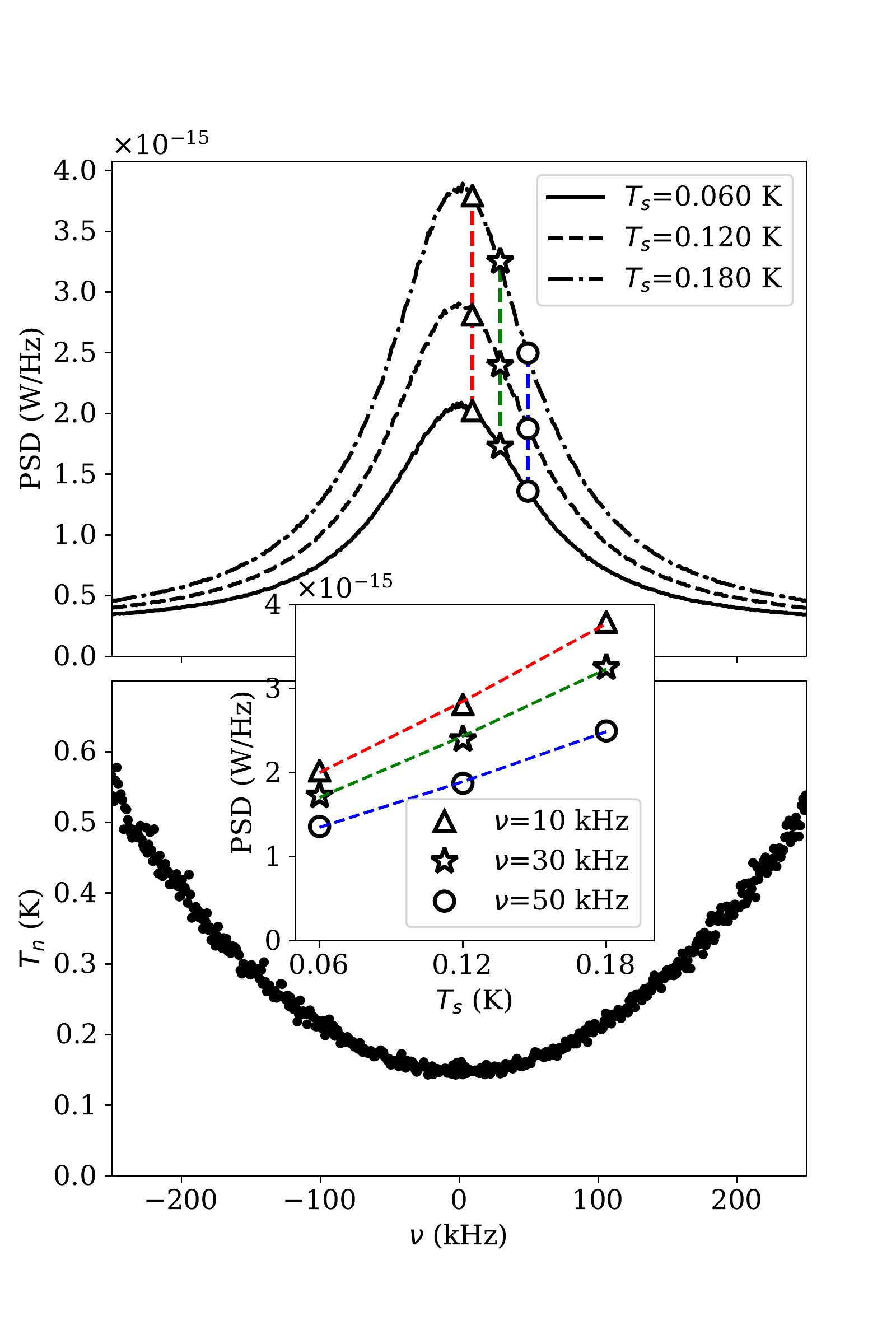}
    \caption{The upper plot shows the set of power spectra obtained during a noise temperature
    measurement performed for a tuning at $ f_r =
    \SI{2.305}{\giga\hertz}$ with $ f_p = 2f_r $.  The offset $ \nu $ is defined
    as $ \nu = f - f_r $ where $ f $ is the center of the frequency bin at which
    the power was measured using the spectrum analyzer.  $ T_s $ is the
    temperature of the noise source.  The inset shows three vertical slices which
    were fit with \cref{eq:ntfitform}. The lower plot shows the estimated noise temperature of the whole
    chain as a function of $ \nu $.}\label{fig:ntsingle}
\end{figure}

To investigate noise temperature, a methodology similar to the well-known
Y-factor method\cite{Engen1970} was used.  A \SI{50}{\ohm} cryogenic microwave
terminator was used as the noise source.  A bias-tee was attached in front for
improved thermalization of its inner conductor.  These two
components were fixed onto a gold-plated copper plate along with a ruthenium
oxide temperature sensor and a \SI{100}{\ohm} resistor functioning as a heater.
This plate was then fixed onto the MC plate so that the dominant
thermalization was through a thin copper wire attached to the MC 
plate.  The noise source was connected to the switch input using a
superconducting coaxial cable, which provides thermal
isolation while minimizing losses.  Using a PID controller, the terminator
temperature could be adjusted from \SI{50}{\milli\kelvin} to \SI{1}{\kelvin}
without affecting the MC plate temperature.  The noise power generated by
the noise source was measured using a spectrum analyzer (SA) with
\SI{1}{\kilo\hertz} resolution bandwidth after being amplified by the JPA and
the rest of the signal detection chain.  The power spectra were recorded at noise source
temperatures ($ T_s $) of \SIlist[list-units=single]{60;120;180}{\milli\kelvin}.  The
power values were converted into power spectral densities (PSD) by dividing
them with the noise bandwidth corresponding to the SA settings used.  Before
each PSD measurement, the JPA gain and passive resonance were measured.  From
these measurements, it was concluded that there were neither gain changes nor
resonance shifts.   From the obtained PSD values $ S(T_s) $, a fit was done to a
function of the following form independently for each frequency bin~(see
~\cref{fig:ntsingle})~:
\begin{align}\label{eq:ntfitform}
    S(T_s) = (2 G_J - 1) \frac{G_L G_\mathrm{tot}}{G_J} (S_n(T_s) + r k_B T_n + \gamma) 
\end{align}

where $ S_n(T_s) $ is the noise PSD of the source, $ G_\mathrm{tot} $ is the total
gain seen from the reference plane, $ G_L $ is the loss factor between the
\SI{50}{\ohm} terminator and the reference plane and $ T_n $ is the noise
temperature.  The reference plane is at the end of the superconducting cable
connected to the noise source (see \cref{fig:setup}).
Here, $ r $ and $ \gamma $ are factors that are explained in
\cref{appendix:ntest}.

\begin{figure}[h]
    \centering
    \includegraphics[width=\linewidth]{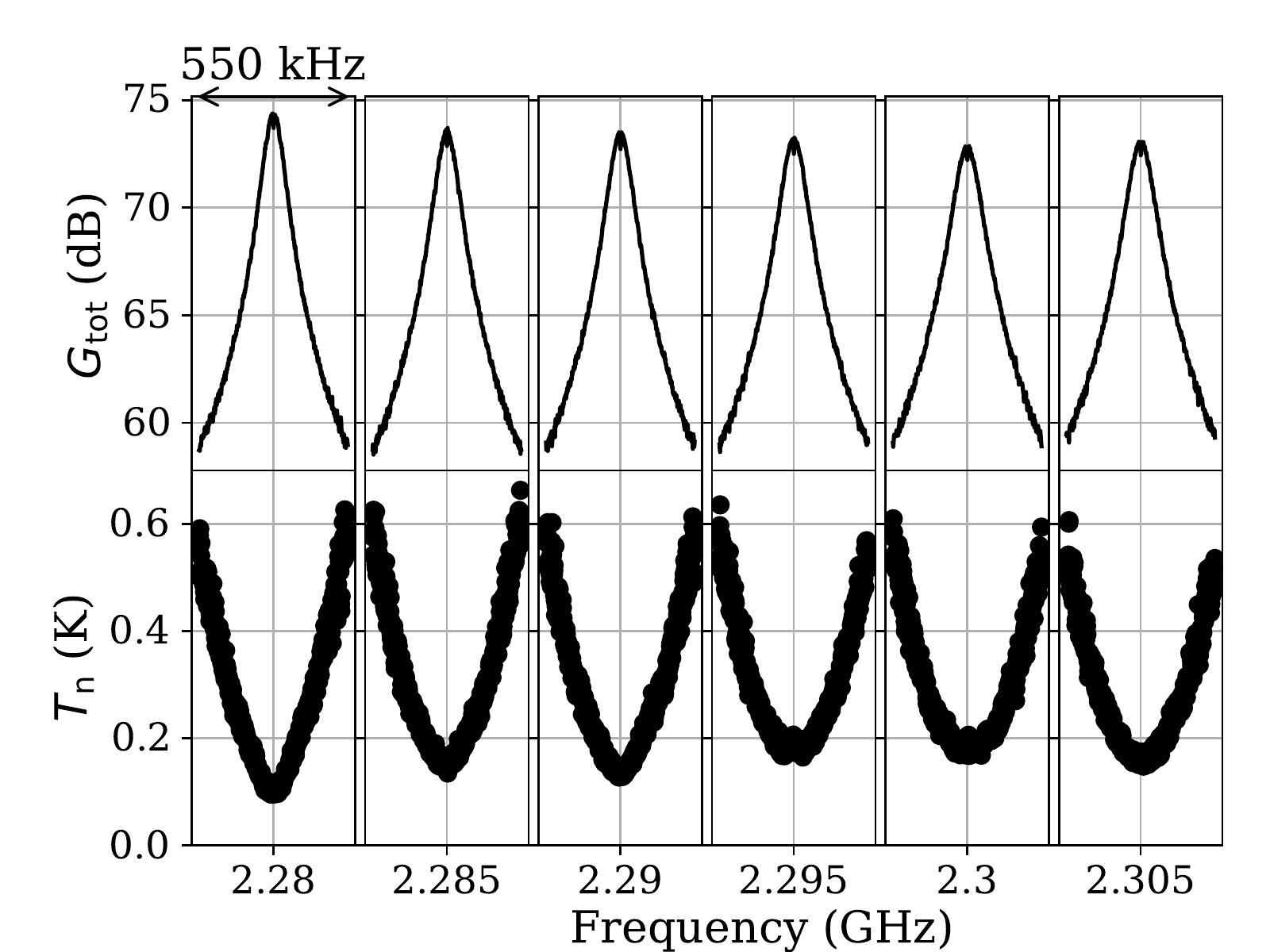}
    \caption{Total gain and the noise temperature of the whole chain for 6
    tuning points with \SI[separate-uncertainty=true]{19.3 \pm{}0.5}{\decibel} JPA
    gain.  Both quantities were estimated from noise temperature measurements.
    The small variations along tuning frequencies were mainly attributed to the
    losses due to the microwave components before the JPA.}\label{fig:ntsweep}
\end{figure}

Since the amplifier needs to be tuned along with the cavity during the axion
experiment, the noise temperature was investigated at different frequencies.
The measurements were done in \SI{5}{\mega\hertz} steps from
\SIrange{2.28}{2.305}{\giga\hertz}.  At each step, the pump power and resonance
frequency were tuned such that the JPA gain was about $ 20 $ \si{\decibel}.
From these measurements (\cref{fig:ntsweep}) a minimum noise temperature of
\SI{120}{\milli\kelvin} was observed at \SI{2.28}{\giga\hertz}.

\begin{figure}[h]
    \centering
    \includegraphics[width=\linewidth]{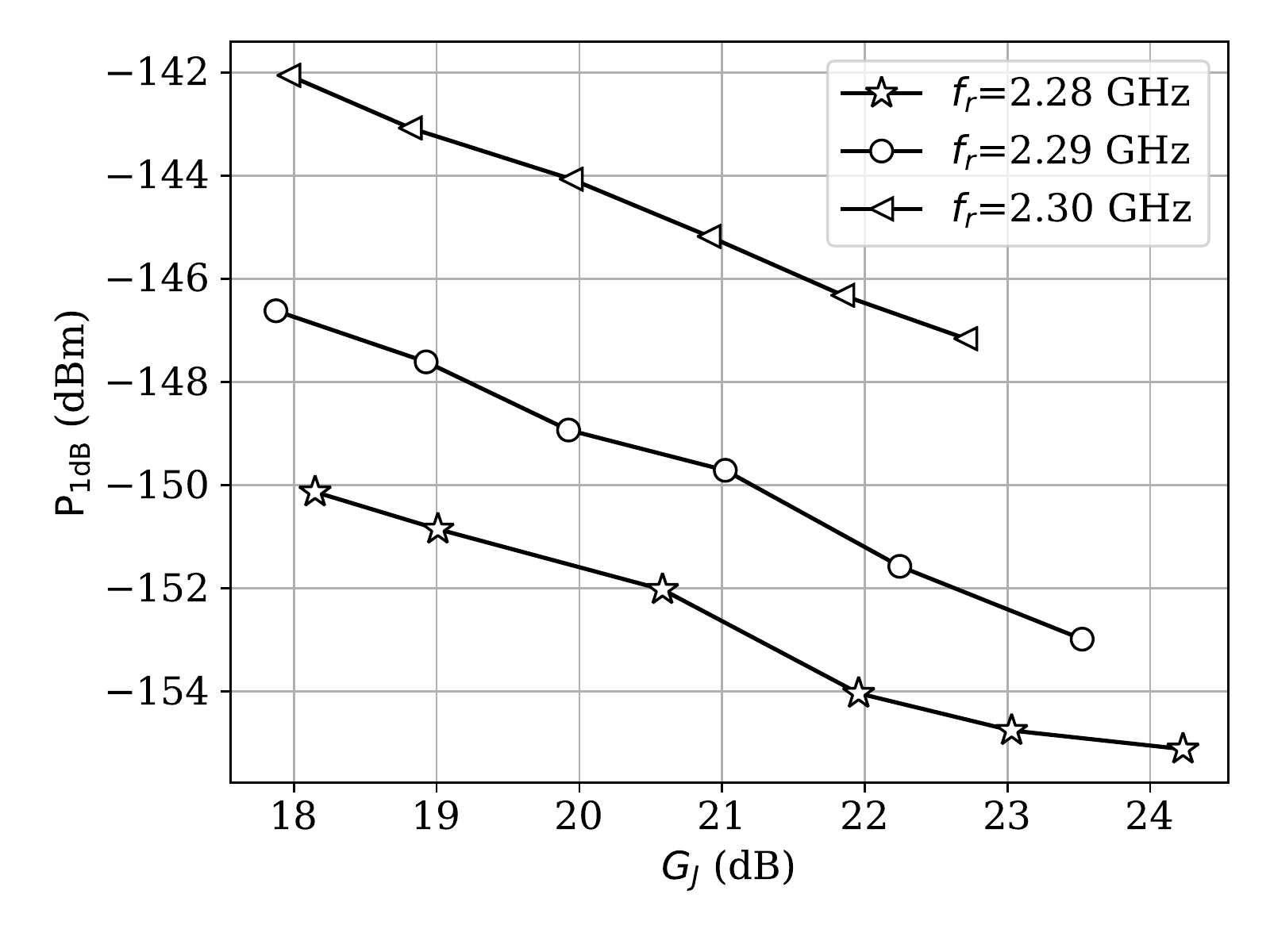}
    \caption{Saturation measurements for three different $ f_r $.  Each
    measurement was done by sweeping the signal powers from the VNA and observing
    at which input power the gain reduces by 1 dB.  The horizontal axis
    corresponds to the unsaturated gain measured with the lowest signal power
    available from VNA.}\label{fig:p1db}
\end{figure}

Another important characteristic is the saturation that occurs when a narrowband signal is
applied.  For stable and predictable operation the JPA must be operated away
from the effects of saturation.  A common way to quantify the saturation of an
amplifier is to determine the input power at which the gain is reduced by
\SI{1}{\decibel} ($ P_{1\mathrm{dB}}$).  The $ P_{1\mathrm{dB}}$ was measured at $ \delta = 0 $
for different frequencies and different pump powers corresponding to different
gains.  It is evident from the results (see \cref{fig:p1db}) that an
axion-like signal with an expected power of \SI{-180}{\dBm} is far from
saturating the device.  While saturation from narrowband signals is avoidable
to a certain extent, it was observed that thermal noise at the input can also
saturate and alter the behavior of the device.  For frequencies below
\SI{2.28}{\giga\hertz} with gains above \SI{23}{\decibel}, the device started
showing saturated behavior with thermal noise when the noise source
temperature was raised above \SI{120}{\milli\kelvin}, which was done to measure noise
temperature.  While this does not necessarily mean that the device is unusable
below these frequencies, it renders the direct measurement of the noise
temperature using a noise source unreliable for these frequency and gain
regions.

\section{Conclusion}\label{sec:conclusion}
In conclusion, a flux-driven JPA, tunable in the range
\SIrange{2.2}{2.305}{\giga\hertz} was demonstrated and determined to be operational for use in
axion search experiments.   The added noise temperatures of the receiver chain
were measured using a noise source at a location as close as possible to the
origin of the axion signal.   With an added noise temperature of 
\SI{120}{\milli\kelvin} the system was shown to reach $ T_\mathrm{sys} \approx
1.7 T_\mathrm{SQL} $.  This is the first record of $ T_\mathrm{sys} $ below $ 2
T_\mathrm{SQL} $ for an axion haloscope setup operating below
\SI{10}{\giga\hertz}. The saturation input power for the JPA was observed
to be more than adequate for an axion-like signal.  Currently, the tested JPA
is being used as part of a KSVZ\cite{Kim1979,Shifman1980} sensitive axion search experiment at the Center
for Axion and Precision Physics Research (CAPP).  The system is taking physics data with a 
scanning speed that has been improved more than an order of magnitude.  We expect that
further optimization of the JPA design could result in improved instantaneous
bandwidth and tuning range.
  
This work was supported by the Institute for Basic Science
(IBS-R017--D1--2021--a00) and JST ERATO (Grant No. JPMJER1601).  A. F. van Loo
is supported by a JSPS postdoctoral fellowship.




%
%

%


\appendix
\section{Noise Temperature Estimation}\label{appendix:ntest}

\begin{figure}[h]
    \centering
    \includegraphics[width=0.95\linewidth]{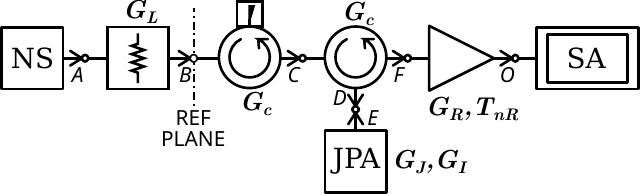}
    \caption{The simplified model used for the noise temperature estimations
        conducted in this work.  Bold letters denote the power gains of
        components.  The reference plane marks the input of the detector chain.
        Arrows denote the flow of power entering the nodes shown with a small
        circle.  $ G_L $ is a composite gain factor for everything between the
        noise source and the reference plane.  $ G_c $ is the circulator gain
        factor.  $ G_J $ is the signal and $ G_I $ is the idler gain for the
        JPA\@.  The amplifier gain $ G_R $ and noise temperature $T_{nR}$
        contain the effects of all elements after the last circulator,
    including SA noise. For simplicity, circulators are assumed to have
complete rotational symmetry with respect to their ports and to be completely
identical to each other.}\label{fig:ntmodel}
\end{figure}

The output PSD from a component with its input connected to a matched source can be written as~:
\begin{align}
    S_\mathrm{O} = G
    S_{\mathrm{in}} + S_\mathrm{added}
\end{align}
where $S_\mathrm{in}$ is the source PSD, $G$ is the power gain of the
component, and $ S_\mathrm{added} $ is the noise added by it.  The
noise temperature ($ T_n $) is a measure of the added noise at the output of a
component.  By convention, it is defined as if it is for noise entering the
device itself: $ T_n = S_\mathrm{added}/(k_B G) $.  The entire detection
chain (see \cref{fig:ntmodel}), from the reference plane to the spectrum
analyzer, can be described as a single composite component with $ G =
G_\mathrm{tot} $ and noise temperature $ T_n $. 

The noise temperature can be defined for a situation similar to the experimental
one where a narrowband axion signal with power $ A $ is present.  This signal
enters the chain from a source connected to the reference plane.  Assuming the
source is thermalized to the MC plate with the temperature $ T_f $, then
the defining relation for $ T_n $ can be written as~:

\begin{align}
\label{eq:tndef}
S_O = G_\mathrm{tot} (\underbrace{A \delta(f-f_s) + S_n(T_f)}_{S_\mathrm{in}} + k_B T_n)
\end{align}

where $ S_O $ is the PSD at the output, $ G_\mathrm{tot} $ is the total power
gain for the signal from the reference plane, $ S_n $ is the noise coming from
the source itself.  The main idea here is that if one has a reliable estimate
of $ T_n $, and understands the source environment well ($ S_n(T_f) $), it is
straightforward to estimate $ A $ without the precise knowledge of $
G_\mathrm{tot} $.  This is possible since $ S_O $ can be easily measured at two
frequencies $ f_s $ and $ f_s' $ using a spectrum analyzer.  Provided that $ |
f_s - f_s' | $ is small enough so that $ T_n $ is approximately the same for
both frequencies, $ A $ can be estimated from these two measurements.  This
approach forms the basis of the analysis methods applied in axion dark matter
search experiments\cite{Lee2020, Jeong2020, Asztalos, Brubaker2017}.

The detection chain consists of passive components, the JPA and the HEMT
amplifiers.  Each one of these adds noise in a different way.  A passive
component at physical temperature $ T_f $ has $ S_\mathrm{added} = (1-G)
S_n(T_f) $.  The HEMT amplifier noise is usually estimated from measurements.  The JPA
adds noise by two main mechanisms.  The first one is by amplifying the input
noise at the idler mode onto the signal mode.  The second one is via the losses
or other dissipation mechanisms inside or before the sample.  Ideally, the
latter can be made zero, whereas the former will approach to the half-photon
added noise in the limit of a \SI{0}{\kelvin} bath temperature.

Using the model shown in \cref{fig:ntmodel}, it is straightforward to write a
relation for the output PSD\@.  For clarity, the explicit frequency dependence of
the thermal noise $ S_n $ and of the gains will be omitted.  Also, the
approximation $ S_n(f, T) \approx S_n(f_p - f, T)$ will be denoted with the
shorthand $ S_{nf} = S_n(f, T_f)$.  This approximation has less than 30 ppm
error given that $ | 2f - f_p | < \SI{100}{\kilo\hertz}$.  Note that
\SI{100}{\kilo\hertz} is the typical bandwidth for the JPA tested in this work.
Furthermore, the transmission characteristics of the microwave components will
be assumed to not vary on a scale of \SI{100}{\kilo\hertz}.  Using the gain
symbols for components as shown in \cref{fig:ntmodel}, the power flow at each
node in terms of their PSD is written as~:

\begin{align}
    \begin{split}
S_B &= G_L S_A + (1- G_L) S_{nf} \\
S_C &= G_c S_B + (1- G_c) S_{nf} \\ 
S_D &= G_c S_C + (1- G_c) S_{nf} \\ 
S_E &= G_J S_D + G_I S_D + G_J S_j \\ 
S_F &= G_c S_E + (1- G_c) S_{nf} \\ 
S_O &= G_R(S_F + k_B T_{nR})
\end{split}\label{eq:pflow}
\end{align}

The idler gain is denoted by $ G_I $, and is substituted using $
G_I = G_J - 1$\cite{Yurke1989} in the following derivations.  As shown in
\cref{eq:pflow}, the idler contribution to the noise appears as $ G_I S_D $.
The symbol $ S_j $ denotes an unknown noise density added at the JPA stage
which does not contribute to the quantum limit but rather contains losses or
other mechanisms of stationary noise.  Note that the noise propagating back 
from the later stages is also included in $ S_j $.  The output $ S_O $ can be
written for two cases.  In the first case, the noise source is operational at
temperature $ T_s $, and in the second case, a signal source at temperature $
T_f $ is connected to the reference plane.  The former case describes the 
measurement situation, whereas the latter case is only used to define $
T_n $ in terms of the parameters in the model.  For the first case, \idest{} $
S_A = S_n(T_s) $, the output PSD can be written as~:

\begin{align}
	\label{eq:sontestrelation}
    S_O^{(1)} &= \underbrace{G_R G_c^3 G_L (2G_J - 1)}_{G_{\mathrm{noise}}}
        \left[ S_n(T_s) + S_\alpha \right] \\
	S_\alpha &= \lambda^{(1)} S_{nf} + 
        \frac{G_J S_j}{(2 G_J -1)G_c^2 G_L} +
        \left. \frac{k_B T_{nR}}{G_c^3 G_L (2 G_J-1)} \right.
        \label{eq:salphadef}\\
    \lambda^{(1)} &= \beta_l + \frac{\beta_c}{G_L} + \frac{\beta_c}{G_c G_L} + \frac{\beta_c}{G_c^2 G_L}\\
    \beta_{\plchldr} &\equiv \frac{1}{G_{\plchldr}} - 1
\end{align}

The output for the second case, where $ S_B = A \delta(f - f_s) + S_{nf} $, is
written as~:

\begin{align}
\label{eq:sontrelation}
S_O^{(2)} &= \underbrace{G_J G_c^3 G_R}_{G_{\mathrm{tot}}} \left[S_B + k_B T_n \right]\\
k_B T_n &= \lambda^{(2)} S_{nf}
	+ \frac{S_j}{G_c^2} 
    + k_B \frac{T_{nR}}{G_c^3 G_J} \label{eq:tndef2}\\
\lambda^{(2)} &= \left[
	\frac{G_J-1}{G_J} 
	+ \left(\beta_c + \frac{\beta_c}{G_c}\right)\frac{2G_J-1}{G_J}
	+ \frac{\beta_c}{G_c^2} \right]\label{eq:lambda2}
\end{align}

Here, the unknowns are $ S_j $, $ T_{nR} $ and $ G_R $.  It is clear from
\cref{eq:tndef2,eq:lambda2} that $ T_n $ approaches $ T_Q $ as expected in the
limits of $ G_J \gg 1$, $ G_c\rightarrow 0$, and $ G_L\rightarrow
0$.  Using \cref{eq:sontestrelation,eq:salphadef,eq:tndef2} $, S_O^{(1)} $ can
be rewritten as~:

\begin{align}
    \label{eq:ntfitrel}
    \begin{split}
        S_O^{(1)} &= \frac{G_{\mathrm{tot}}}{r}(S_n(T_s) + r k_B T_n + \gamma) \\
    r &= \frac{G_J}{G_L(2 G_J - 1)}\\
    \gamma &= \left(\lambda^{(1)}-r\lambda^{(2)} \right)S_{nf}
    \end{split}
\end{align}

This relation is used to perform a fit with $ G_{\mathrm{tot}}$ and $ T_n $
as the fit parameters.  For the estimations, the parameters $ G_L $ and $ G_c $ were
taken as \SI{-0.05}{\decibel}, and \SI{-0.4}{\decibel} respectively.  Some
typical values for $ r $ and $ \gamma/k_B $ can be found in \cref{tab:rgamma}.

\begin{table}[ht]
    \caption{\label{tab:rgamma}Calculated $r$ and $\gamma /k_B$ for
    typical component losses and $T_f=\SI{50}{\milli\kelvin}$.}
    \begin{tabular}{@{}ccccc}
        \toprule
        $ G_J (\si{\decibel}) $ & $ G_L(\si{\decibel}) $ & $G_c(\si{\decibel})$ & $ r $ & $ \gamma/k_B(\si{\milli\kelvin}) $\\
        \midrule
        15 & 0 & -0.3 & 0.508 & -31.1 \\
        15 & -0.2 & -0.5 & 0.532 & -26.8 \\
        20 & 0 & -0.3 & 0.503 & -31.4 \\
        20 & -0.2 & -0.5 & 0.526 & -27.1 \\
        19 & -0.05 & -0.4 & 0.509 & -29.8 \\
        \bottomrule
    \end{tabular}
\end{table}
\bibliography{fdjpa-axion_arxiv}

\end{document}